\documentclass[a4paper,onecolumn,11pt]{article}
\usepackage[dvips]{graphicx}
\usepackage{natbib}
\usepackage{url}

\title{Building a Model Astrolabe}
\author{Dominic Ford}
\date{\footnotesize Published in the Journal of the British Astronomical Association, February 2012. \\ Supplementary materials available for download: \\ \url{http://dcford.org.uk/astrolabe/index.html} \\ Submitted: October 2010. Accepted: December 2010.}
\begin{document}
\maketitle

\begin{abstract}
This paper presents a hands-on introduction to the medieval astrolabe, based
around a working model which can be constructed from photocopies of the
supplied figures. As well as describing how to assemble the model, I also
provide a brief explanation of how each of its various parts might be used. The
printed version of this paper includes only the parts needed to build a single
model prepared for use at latitudes around 52$^\circ$N, but an accompanying
electronic file archive includes equivalent images which can be used to build
models prepared for use at any other latitude.  The vector graphics scripts
used to generate the models are also available for download, allowing
customised astrolabes to be made.
\end{abstract}

\section*{Introduction}

For nearly two thousand years, from the time of Hipparchus ({\it c}.\ 190--120
BCE) until the turn of the seventeenth century, the astrolabe was the most
sophisticated astronomical instrument in widespread use. Yet today this complex
instrument is rarely seen, and those interested in learning about it may even
have some difficulty finding a specimen to play with. Ornately carved brass
reproductions are available from several telescope dealers, but with
substantial price tags attached. These price tags are historically authentic:
medieval astrolabes were often made from high-cost materials and intricately
decorated, becoming expensive items of beauty as well as practical observing
instruments. But for the amateur astronomer who is looking for a toy with which
to muse over past observing practice, a simpler alternative may be preferable.

In 1975--1976, Sigmund Eisner contributed a series of three papers
\citep{r1,r2,r3} to the {\it Journal of the British Astronomical Association}
entitled {\it Building Chaucer's Astrolabe}. In them, he described how a
cardboard astrolabe might be built, using as a model the instrument described
by the English poet Geoffrey Chaucer ({\it c}.\ 1343--1400) in his {\it
Treatise on the Astrolabe} \citep{r4}. Though the task described is a
time-consuming exercise in geometry, it gives a rewarding insight, not only
into how astrolabes were used, but also into their detailed construction and
workings.  With the advent of computerised vector graphics, it has become
possible to automate much of the delicate geometric construction work that
Eisner describes. This paper describes the result of implementing a slightly
modified version of Eisner's instructions in a computerised vector graphics
scripting language called {\it
PyXPlot}\footnote{\url{http://www.pyxplot.org.uk/} (available for Linux/MacOS X
only)}, the interpreter for which is available for free download under the GNU
General Public License
(GPL)\footnote{\url{http://www.gnu.org/licenses/gpl-2.0.html}}. The various
parts of the resulting model are shown on subsequent pages; a later section
will describe how they should be cut out and assembled. As an alternative to
working with photocopies of the printed version of this paper, these figures
are also available in computer printable PDF format from the electronic file
archive which accompanies this paper; this can be downloaded from the author's
website at \url{http://dcford.org.uk/astrolabe/index.html}.

As well as the figures printed here, which are prepared specifically for use at
latitudes around 52$^\circ$N, these archives also contain equivalent figures
which may be used to build astrolabes prepared for use at other latitudes. In
addition, they include the full {\it PyXPlot} scripts used to generate the
model, which may be distributed freely under the GPL or modified to produce
custom astrolabes.

\section*{The astrolabe}

An astrolabe typically takes the form of a disc, often made of wood or brass,
around 10--20 centimetres in diameter and a few millimetres thick. An eyelet
protrudes from one side of the disc, through which a ring is connected as a
handle. The body of this disc is called the {\it mother}, and one of its sides
is designated as its front and the other as its back. Two freely rotating
pointers are mounted on a central pivot, one on each side of the mother. The
pointer on the back of the mother, known as the {\it alidade}, is used as a
line along which to sight celestial or terrestrial objects when making
approximate measurements of their altitudes. The front side of the instrument
as a whole can be roughly described as a more sophisticated sibling of the
modern planisphere, providing a way of predicting the altitudes and azimuths of
celestial objects at any given time.

\begin{figure}
\centerline{\includegraphics{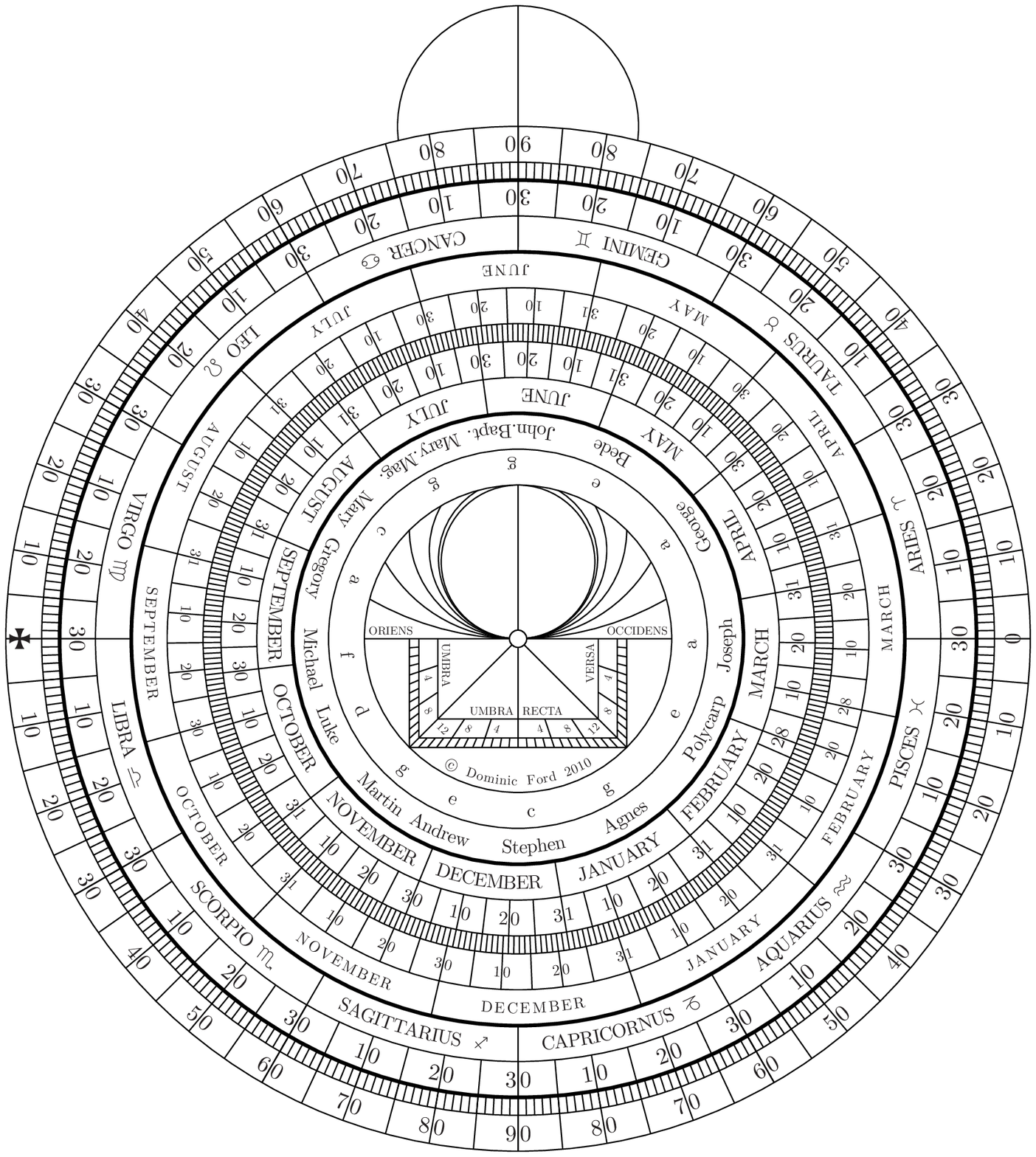}}
\caption{The back of the mother of the astrolabe.}
\label{fig:1}
\end{figure}

Beyond this rather loose description, astrolabes were historically built to
many diverse designs. The emergence of the astrolabe as a single instrument
began with the bringing together of two forerunners in Greece in the second
century BC: the dioptra -- an instrument for sighting the altitudes of
celestial bodies -- and planispheric projections which could be used to
represent the celestial sphere on a flat surface. Together, they formed a
single instrument which could simultaneously measure or predict the positions
of celestial objects as required, becoming what might be described as an
analogue celestial calculator. This powerful hybrid instrument spread to the
Byzantine, Islamic and Persian worlds over subsequent centuries, evolving
variously along the way. Some of the products of this evolution -- for example,
linear and spherical astrolabes -- do not even fit within the deliberately loose
description above, though they retain a common {\it raison d'\^etre}. Another
product, the mariner's astrolabe, emerged as a similar but distinct instrument,
simplified and optimised for use on the deck of a rolling ship in the
determination of latitude at sea. The subject of this paper is the
astronomical astrolabe, which retained its same essential layout.

It is beyond the scope of this paper to discuss the historical development of
the astrolabe any further, or to present a study of its historical use. The
interested reader may find a fuller account of these subjects in John North's
comprehensive history of astronomy, {\it Cosmos} \citep{r7}.  My aim in
presenting a single design of model astrolabe here is rather to provide a
hands-on introduction to how a modern amateur astronomer might use such an
instrument to make similar observations to those for which the astrolabe was
historically used.  The particular design of astrolabe which I present is
loosely based upon one described in the 14th century by Chaucer, a time when
astrolabes had recently arrived in Western Europe, and specifically Britain,
through contact between Christian and Islamic scholars in Spain.

Since my aim is to convey the principles of the instrument rather than its
precise historical appearance, two parts of the astrolabe have been slightly
modernised to make them more immediately usable today. The calendar appearing
on the astrolabe described by Chaucer has been updated from the Julian calendar
to the Gregorian calendar, though the original is retained alongside its modern
counterpart. More significantly, the star chart, known as the rete, has been
transferred from brass latticework to transparent plastic, as will be described
later. These changes notwithstanding, the principles of the observations and
calculations which I describe are similar to and largely based upon those
described by Chaucer and other historical sources, though medieval users would
often have had astrological cosmologies in mind which are foreign to modern
astronomers and are not described here.

\section*{Constructing a model astrolabe}

\begin{figure}
\centerline{\includegraphics{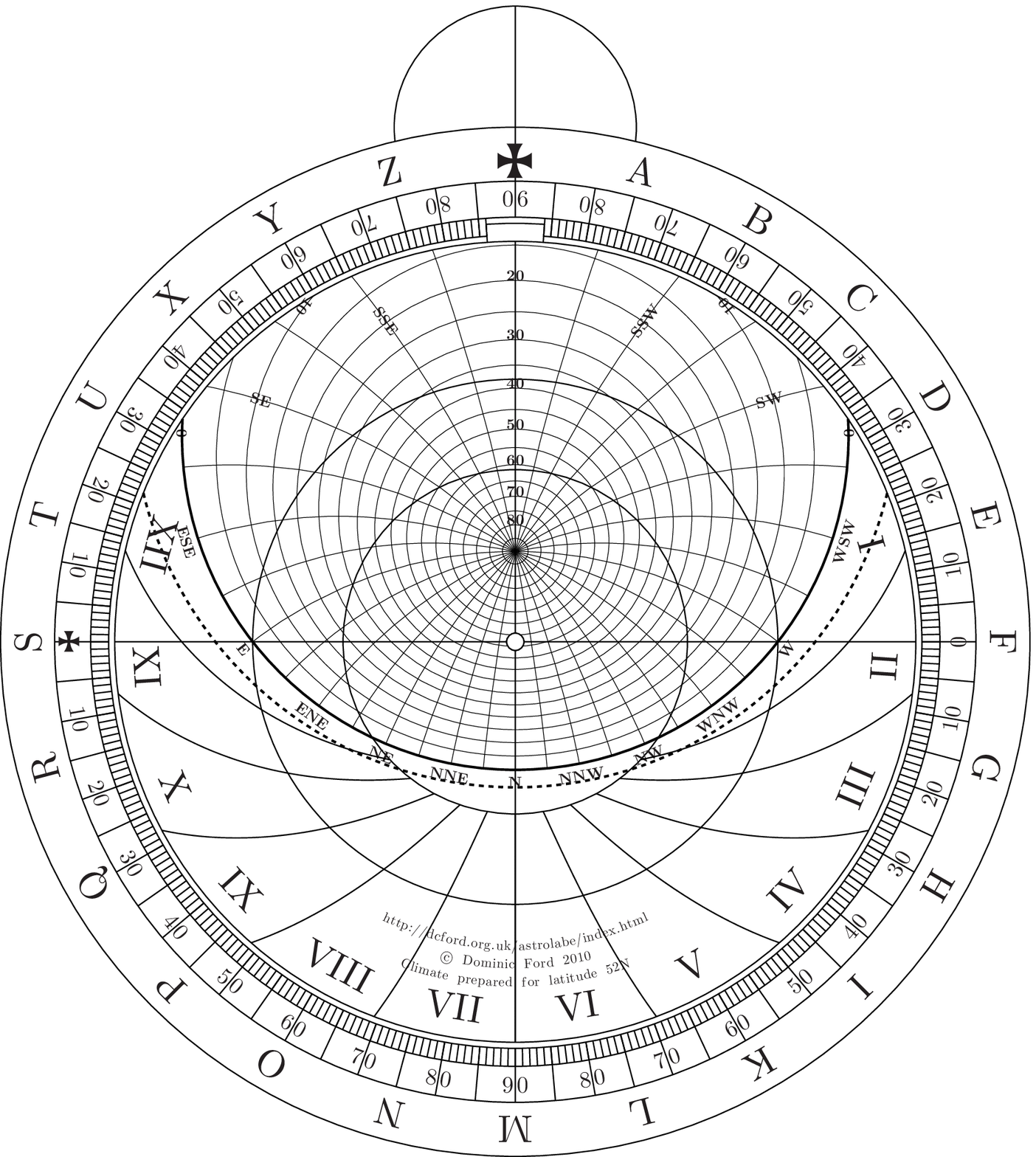}}
\caption{The front of the mother of the astrolabe, with combined climate
prepared for a latitude of 52$^\circ$N. Should a climate for a different
latitude be required, the electronic file archive which accompanies this paper
should be downloaded. This includes separate images of the front of the mother
-- the outer ring of the above image -- and of climates for any latitude on the
Earth at 5$^\circ$ intervals.}
\label{fig:2}
\end{figure}

\begin{figure}
\centerline{\includegraphics{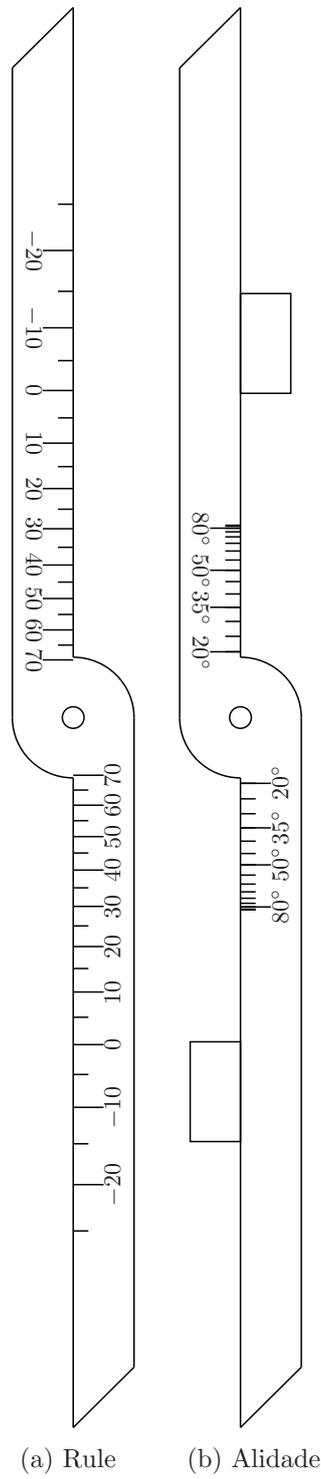}}
\caption{Left: The rule, which should be mounted on the front of
the astrolabe. Right: The alidade, which should be mounted on the
back of the astrolabe.}
\label{fig:3}
\end{figure}

\begin{figure}
\centerline{\includegraphics{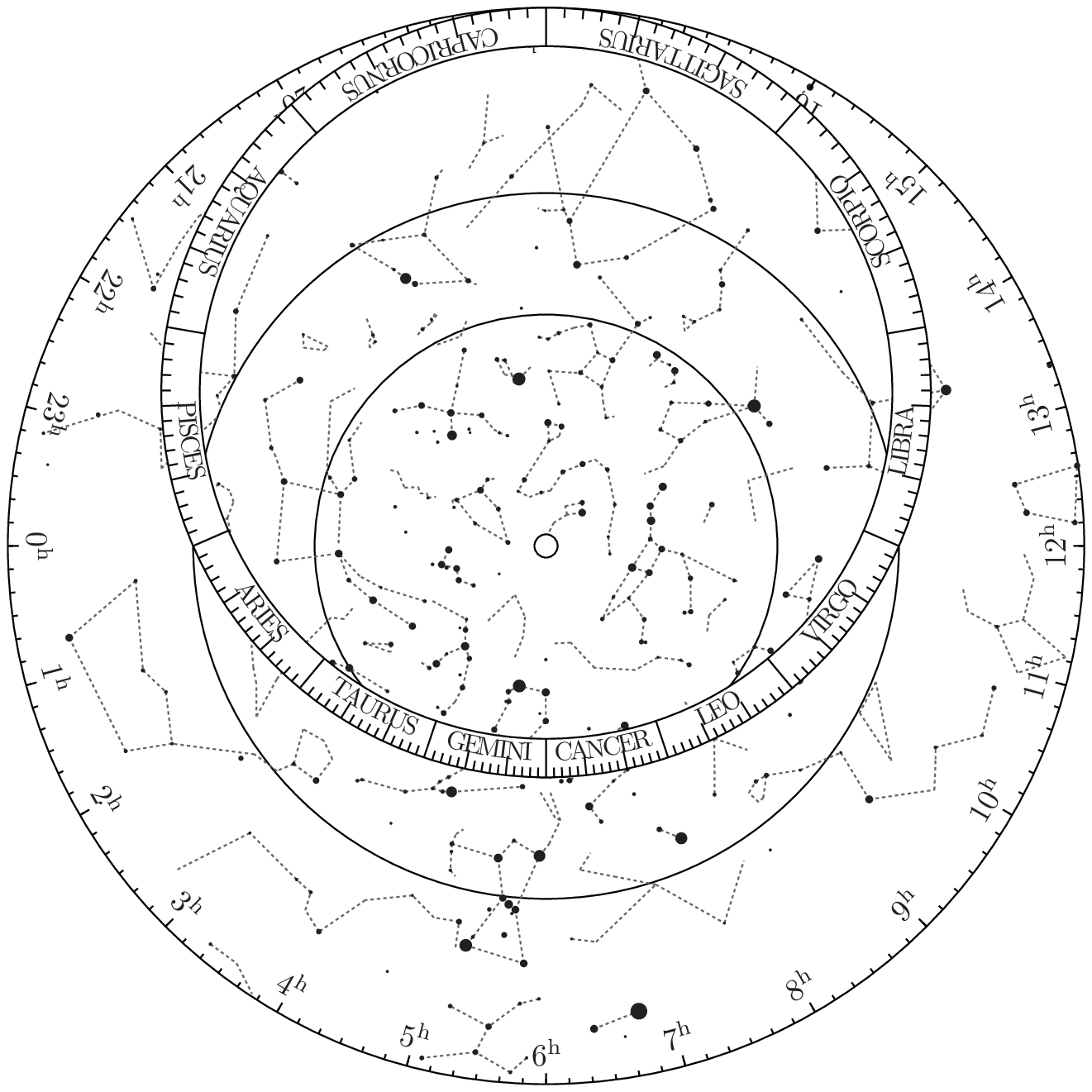}}
\caption{The rete of the astrolabe, showing the stars of the northern sky. This should be printed onto a piece of transparent paper; most
stationers should be able to provide acetate sheets for use on overhead projectors, which are ideal for this purpose. Should a southern-hemisphere
astrolabe be required, the electronic file archive which accompanies this paper should be downloaded.}
\label{fig:4}
\end{figure}

To assemble the model astrolabe presented in this paper, Figures~\ref{fig:1},
\ref{fig:2} and~\ref{fig:3} should be photocopied onto paper, or preferably
onto thin card. Figure~\ref{fig:4} should be photocopied onto a sheet of
transparent plastic; acetate sheets, widely sold for printing overhead
projector slides, are ideal for this.  Alternatively, all of the figures in
this paper may be obtained in PDF format from the accompanying electronic file
archive.  Some parts of the astrolabe are drawn with a particular latitude in
mind and space permits only a single astrolabe designed for use at a latitude
of 52$^\circ$N to be included here.  However, the PDF versions of these figures
are accompanied by alternatives which can be used to make model astrolabes for
use at any latitude between 80$^\circ$N and 80$^\circ$S, spaced at 5$^\circ$
intervals.

Once the components of the astrolabe have been printed and cut out,
Figures~\ref{fig:1} and~\ref{fig:2} -- the back and front of the mother --
should be glued rigidly back-to-back, perhaps sandwiching a piece of rigid
card. Figure~\ref{fig:4} -- the rete -- should then be placed over the front of
the mother. From Figure~\ref{fig:3}, the rule, on the left, should be placed
over the rete on the front of the astrolabe, and the alidade, on the right,
should be placed over the back of the astrolabe. The two rectangular tabs on
the alidade should be folded out to form a sight for measuring the altitudes of
objects.  The holes at the centres of each of the components should be cut out,
and a split-pin paper fastener used to fasten the components together, whilst
still leaving them free to rotate. A second hole should be cut into the eyelet
protruding from the top of the mother, allowing the whole instrument to be
suspended from a ring or piece of thread.

The following sections describe in turn how each part of this model astrolabe
may be used.

\section*{The back of the mother}

Figure~\ref{fig:1} shows the features marked on the back of the mother of the
astrolabe. Over the top of this, the alidade, on the right side of
Figure~\ref{fig:3}, is allowed to freely rotate about a central pivot, acting
as a measuring ruler. Two tabs fold out of the edges of the alidade to act as a
sight. To measure the altitude of a star, the astrolabe is suspended freely at
eye level, traditionally from a ring hooked over the user's right thumb, and
the alidade is rotated until both tabs appear in line with the star. The
outermost scale on the astrolabe then indicates the star's altitude.

The remaining circles on the back of the mother are a calendar, referenced
relative to the Sun's annual motion along the ecliptic, such that each degree
of the circular scale represents the time taken for the Sun to move one degree
across the celestial sphere. The ecliptic circle is divided into twelve equal
30$^\circ$ portions, and each portion designated as a zodiacal constellation.
The first of these, Aries, is defined to start at the vernal equinox, the point
on the sky where the Sun's path crosses the equator in March. It should be
noted that these ancient zodiacal constellations bear little relation to the
88~modern constellations known to astronomers today; the latter are a much
newer creation which were only finalised by the International Astronomical
Union (IAU) as a definitive list in 1922 and given rigid boundaries in 1930.

A scale divides the year into twelve parts as the Sun passes through each of
these zodiacal constellations, and within this, a calendar provides a
conversion to the familiar system of days and months. Together, these scales
can be used for determining the celestial position of the Sun on any given
date, which, as we shall see, needs to be known when the front side of the
astrolabe is used. Following Eisner, two calendars are shown. The outer
calendar, in small typeface, is computed for 1394, around the time of Chaucer's
composition, using solar longitude data published by \cite{r8,r9}. This calendar
should be used when following the example calculations described by Chaucer.
The inner calendar is computed for 1974 -- taken as the present day -- using
corresponding data published by \cite{r10} and should be used for modern
calculations.

These two calendars are offset by nine days relative to each other, which can
be explained by the calendrical reforms which took place between 1394 and 1974.
In 1394, the Julian calendar had been in use in Britain since Roman times.
However, because it gave the year an average length of 365.25~days, some
11~minutes longer than its true length, the date of the vernal equinox drifted
by around a day every 128~years.  Though the vernal equinox had occurred around
March~21 at the time of the Council of Nicaea in 325, by 1394 it was falling on
March~12. This was a serious problem for the Catholic Church, which used the
date of the equinox in calculating the date of Easter: should the traditional
Nicene date be used, or that when the equinox was actually observed?

Primarily to resolve this confusion, Pope Gregory XIII reformed the calendar
according to a plan approved by the Council of Trent (1545--1563), removing
centurial leap years such that the average length of the year became
365.2425~days. In addition, the vernal equinox was restored to its traditional
Nicene date by fast-forwarding the calendar by 10~days in 1582, though in
Anglican Britain the new calendar was not adopted until 1752, by which time a
shift of 11 days had become necessary. In contrast to the Julian calendar, the
drift of the Gregorian calendar relative to the seasons is so slight -- 10--20
seconds per year -- that the seasons slip by only one day every few thousand
years, so the calendar presented by Eisner for 1974 remains accurate for modern
use.

The exact choice of the years 1394 and 1974, in each case midway between leap
years, is deliberate. The exact moment of the vernal equinox drifts by a
quarter of a day per year, but shifts back to its original position in the
fourth year. These midpoint years consequently represent an average over the
four-year cycle. In practice, however, this is a detail beyond the accuracy
with which any real measurement can be made with an astrolabe.

In the centre of the back of the mother, a scale shows the dates of a number of
significant saints' days, together with their Sunday letters, acting as
waymarkers through the liturgical year. The choice of saints included here
would vary widely, especially since space only permits the inclusion of one
date every three weeks and local custom would often dictate those included. The
saints included in Figure~\ref{fig:1} were kindly selected by Matthew Smith and
may reflect the bias of a contemporary Anglican. The innermost markings lying
within the circle of saints will be described later.

\section*{The front of the mother}

The front of the mother of the astrolabe is shown in Figure~\ref{fig:2}; the
planispheric starchart in Figure~\ref{fig:3} rotates over the top of the middle
portion of Figure~\ref{fig:2}. This side of the astrolabe is similar to a
modern planisphere, though it is arranged differently and can usually be
adjusted to work at several latitudes, whereas modern planispheres are
typically only designed for one particular latitude.

Around the edge of the mother twenty-four symbols are inscribed, beginning with
a cross and then proceeding through the Roman alphabet. These represent the
twenty-four hours of the day, with the cross representing noon and the letter
`M' midnight. Historically, these characters would have appeared, with the
scale of degrees next to them, on a raised rim around the edge of the mother,
encircling a large well in the middle which Chaucer calls the {\it womb} of the
astrolabe. The circular {\it climate} (sometimes called the {\it plate}
elsewhere) which appears in the centre of Figure 2 would have been a separate
sheet of wood or brass which slotted into the womb. A tab at the top ensures
that the climate is correctly aligned.

The reason for this design is that the positions of the lines drawn on the
climate are latitude dependent, and an astrolabe would typically come with a
supply of different climates for use at different latitudes. Usually the womb
was deep enough that all of the climates could be stacked within for convenient
storage. In Figure 2, for simplicity, only a single climate is provided for
latitude 52$^\circ$N and it is incorporated into the image of the front of the
mother. In the accompanying PDF images, the traditional separation of mother
and climate is observed.

On the rete appears a planispheric projection of the brightest stars in the
northern sky; I have used the Yale Bright Star Catalogue to mark all stars
brighter than fourth magnitude.  The projection used here is the same as is
used on modern planispheres: if a star has right ascension $\alpha$ and
declination $\delta$, then it is plotted at a distance proportional to
$\tan((90^\circ+\delta)/2)$ from the centre and at azimuth $\alpha$. Thus, the
north pole appears at the centre. Three concentric circles are drawn around
this to represent the Tropic of Cancer, the equator, and the Tropic of
Capricorn, which is chosen as the outer edge of the astrolabe. It is not
possible to continue the projection all the way to the south celestial pole, as
this would appear at infinite distance from the centre. The pointer which
rotates over the top of the rete, called the {\it rule} (sometimes the {\it
label} elsewhere), is marked with this mapping between radius and declination,
and can be used to read off the declinations of objects.

Traditionally, the rete would have been made from wood or brass, and arrows
would have pointed to the positions of a small number of stars. Then, as much
of the material of the rete as possible would have been cut out so that the
climate beneath could be seen. Had transparent plastic been available in the
Middle Ages, it would doubtless have been used, and would have allowed many
more stars to be represented. Since the purpose here is not to reproduce a
particular historical astrolabe, but rather to provide a working specimen that
might be used by amateur astronomers today, this is the one component of the
instrument where I have taken the liberty of substantial modernisation. It
should also be noted that the scale of right ascension shown around the edge of
Figure~\ref{fig:4} would not have been present on historical instruments, but
is provided as a navigational aid for the modern astronomer.

The annual path of the Sun across the celestial sphere is marked on the rete by
the outer edge of a circular band on which the positions of the zodiacal
constellations are marked out. As before, each represents an equal 30$^\circ$
portion of the ecliptic, though in the planispheric projection the northern
constellations appear artificially shrunken compared to those in the south. To
find the position of the Sun on any given day, the scale on the back of the
mother is used. For example, on 1st June, the back of the mother tells us that
the Sun has moved approximately 10$^\circ$ through the constellation of Gemini.
Thus, on the rete, we see that the Sun is just a little to the north of
Aldebaran.

The climate sits directly behind the rete and shows the horizon of the visible
sky.  Just as on a modern planisphere, the diurnal passage of the stars is
reproduced by rotating the projection, and so as the rete is rotated clockwise,
stars are seen to rise in the east and set in the west. The cobweb-like grid
that criss-crosses the visible sky shows lines of constant altitude -- called
{\it almucantars} -- and lines of constant azimuth -- called {\it azimuths} --
for determining the approximate alt/az coordinates of stars. Just beneath the
horizon, a dotted line marks the path six degrees beneath the horizon and may
be used to calculate the times of civil twilight, defined to be when the Sun is
between zero and six degrees below the horizon.

As is the case with a planisphere, the projection must be brought into the
correct rotation to represent a particular time and date before it can be used.
On a planisphere, this is usually done by matching the desired time on a
rotating scale to the desired date on a static scale. On an astrolabe, however,
no such scales are provided. Alignment is typically achieved by measuring the
altitude of a reference object -- either the Sun or a star -- using the alidade
and then rotating the rete until its projection lies on the appropriate
almucantar; it is also necessary to have some sense of east and west in order
to know whether to align the object so that it is rising or setting. This means
that, in contrast to the planisphere, the time of day does not need to be
accurately known {\it a priori} to align an astrolabe. The astrolabe can even
be used to determine the time from altitude measurements.

As Chaucer observes, it is best when aligning the rete to select a reference
object which is well away from the meridian. When an object transits, its
altitude is momentarily unchanging, and even the slightest uncertainty in its
altitude leads to a large uncertainty in the time. By contrast, when an object
is in the east or the west, its altitude is changing most rapidly and there is
no such difficulty.

\section*{The unequal hours}

Before the advent of reliable mechanical clocks, it was common to divide each
day not into twenty-four equal hours, but into twelve equal hours of daytime
and twelve equal hours of nighttime. In winter, each hour of nighttime would be
considerably longer than each corresponding hour of daytime, and in summer the
converse would be true. These hours are hence known as {\it unequal hours}. The
next section describes how to align the astrolabe using the equal hours with
which we are now familiar, but I first describe how to use the system of
unequal hours which would have been in widespread use in the Middle Ages.

The area of the climate beneath the horizon is divided into twelve curved
strips. These relate to the fact that the circular path traversed by a star at
any given radius from the centre of the rete -- the radius representing its
declination; see above -- can generally be divided into a portion which lies
above the horizon marked out on the climate, and a portion which lies beneath
the horizon, unless the star is circumpolar. The portion of this circular path
which lies beneath the horizon is divided into twelve equal parts by these
curved strips.

To use these strips to tell the time, it is first necessary to determine the
position of the Sun along the ecliptic using the scale on the back of the
mother, as described previously. The position of the point diametrically
opposite the Sun along the ecliptic -- the antisolar point -- should also be
noted, by looking at the point on the opposite side of the back of the mother.
Turning the astrolabe over, the rete should then be aligned, using a
measurement of the altitude of either the Sun or a known star. Depending on
whether it is day or night, either the antisolar point or the Sun respectively
will be beneath the horizon; at sunset or sunrise, both will be on the horizon.
At nighttime, the number of the strip in which the Sun lies is the hour of the
night; the strips equally divide its path from the point where it sets on the
western horizon to that where it rises on the eastern horizon. Conversely, in
the day, the number of the strip in which the antisolar point lies is the hour
of the day. At sunset and sunrise, the point used to determine the hour
changes, and because the Sun and the antisolar point have opposite
declinations, the astrolabe correctly produces daytime and nighttime hours of
different lengths.

The calculation can also, of course, be done in reverse. To align the astrolabe
to show what the sky would look like at any given hour of the day, the rete
should be turned until the solar or antisolar point is in the correct place
among the lines of the unequal hours.

\section*{The equal hours}

The same alignments can also be made using the system of equal hours with which
we are now more familiar, in which each day is divided into twenty-four hours
of equal length.  To do so, the sequence of twenty-four symbols around the edge
of the mother is used, each of which signifies an hour of the day, with the
cross marking noon and the letter `M' midnight. The procedure is much simpler
than that used to align to the unequal hours: the rule should be rotated to
point to the desired hour around the outer scale, and the rete rotated beneath
it until the Sun lies on the edge of the rule.

One detail is worthy of note here: the twenty-four symbols refer to the hours
of local {\it apparent solar time}, defined such that noon occurs on any given
day when the Sun is at its highest altitude in the sky. This may be found to be
offset relative to civil time for two reasons. First, the observer will, in
general, be some distance east or west of the meridian for which his civil
timezone is defined. Secondly, the speed of the Sun's motion in right ascension
varies over the course of the year such that days in June and December are a
few seconds longer than those in March and September. Civil time is a mean
time, in which this variation is averaged over the course of the year, and as
these seconds accumulate day-by-day, apparent noon can drift up to 16~minutes
either side of midday depending upon the time of year. This offset is given by
the equation of time.

\section*{The unequal hours: Method II}

A second tool for calculating the time, in unequal hours, from the altitude of
the Sun appears in the upper half of the central portion of the back of the
mother. This is a simple but imprecise tool, consisting of six partial arcs of
circles all passing through the centre of the astrolabe.  Before using these,
it is necessary to know the maximum altitude at which the Sun will appear -- at
noon -- on the day of observation. This can be determined using the front of
the mother, once the location of the Sun along the ecliptic has been found. The
answer varies little from one day to the next, and so only needs to be looked
up rather infrequently. The scale of degrees marked along the alidade should
then be studied to find the point on the scale corresponding to the maximum
altitude of the Sun, which we shall call X.

The present altitude of the Sun should then be determined using the alidade.
Keeping the alidade pointing to this altitude, the position of the point X
among the six circular arcs should be determined. The smallest circle is drawn
such that the point X always lies on it at noon. The point X crosses each of
the other circles at hourly intervals, each circle being crossed twice each
day, once as the Sun is rising, and once as it is setting. Thus, the gap
between the zero-altitude line and the largest circle represents the first or
the twelfth hour of the day, and the gap between the two smallest circles
represents either the sixth or the seventh hour; it is necessary to determine
whether the Sun is rising or setting to know which.

\section*{The shadow scale}

Presented with the task of determining the distance to a building of known
height $h$ using an observation of the altitude $\theta$ of its highest point,
modern astronomers would probably turn to trigonometry. Reaching for their
calculators, they would evaluate the tangent of $\theta$, which equals the
ratio of the building's height to its distance. The distance to the building
could be found by evaluating $h\times \tan\theta$. Similarly, the height of a
building of known distance $d$ could be found by evaluating $d/\tan\theta$.

The shadow scale, in the lower half of the central portion of the back of the
mother, is an approximate analogue tool for making such calculations without
the need for a separate lookup table of the tangent function, which is instead
marked out on the astrolabe. The range of altitudes between 0$^\circ$ and
45$^\circ$ is divided into twelve parts, with lines denoting the points where
$\tan\theta$ is $1/12$, $2/12$, \ldots, $12/12$. The line where
$\tan\theta=4/12$, for example, is labeled `4'. The use of twelve as a
denominator here is a good choice -- much better than ten, for example --
because of its having six factors: thus $3/12=1/4$, $4/12=1/3$, etc. Thus, when
a building's highest point is at the altitude labeled `4', its height is four
twelfths -- or one third -- of its distance. At an altitude of 45$^\circ$,
$\tan\theta=1=12/12$. Higher altitudes are similarly labelled with numbers
between~1 and~12, denoting the points where $\tan\theta=12/12$, $12/11$,
\ldots, $12/1$.

\section*{Remarks on extreme astrolabes}

The accompanying electronic file archive includes images of astrolabes designed
for use at any latitude between 80$^\circ$N and 80$^\circ$S, sampled at
5$^\circ$ intervals. However, the planispheric projection used in the star
charts of these astrolabes works best at moderate-to-high latitudes. Since the
sky is only mapped as far south as the Tropic of Capricorn, the southernmost
portion of the sky is not shown for any latitude south of 66$^\circ$N. Thus,
whilst astrolabes are provided for all latitudes, those prepared for equatorial
regions are missing large portions of the visible sky.

South of the equator, matters improve again if the projection is reversed and
the celestial south pole placed at the centre of the rete. The notion of such a
southern-hemisphere astrolabe would have been alien to Chaucer, but I
nonetheless provide them for the curiosity of southern readers. In addition to
the change of rete, one further reversal is required: whereas the retes of
northern astrolabes rotate clockwise as time advances, this is reversed for
southern retes. This is because while the Earth rotates clockwise as seen by an
observer looking down on its north pole, it rotates anticlockwise as seen by an
observer looking down on its south pole. In order that the sequence of Roman
letters around the rim of the front of the mother should still represent
twenty-four equal hours, their direction is reversed on southern astrolabes.

Within the Arctic and Antarctic circles, another problem is encountered: the
system of unequal hours becomes ill-defined since the Sun never sets. In the
climates provided for such latitudes, I have chosen a definition such that
polar days and nights are divided into twelve equal hours between successive
midnights. This definition interfaces smoothly with the days at the beginnings
and ends of the polar days and nights where the Sun just rises or sets, and the
unequal hours of the day or night are infinitesimally short.

\section*{Epilogue: The rise of precision astronomy}

By the turn of the seventeenth century, the astrolabe was starting to be
superseded. In 1576, Tycho Brahe laid the foundation stone of {\it Uraniborg},
a research institute on the small Danish island of Hven \citep{r12,r13}. Over
the following 21~years until its abandonment in 1597, this institute brought
about a revolution in pre-telescopic astronomical instrumentation. Among the
instruments pioneered under Tycho's direction were the sextant -- which allowed
the angular distances between stars to be precisely measured -- and the mural
quadrant -- which allowed the altitudes of transiting stars to be measured with
respect to a plumb line indicating the local vertical. At Uraniborg's
observatory, {\it Stjerneborg}, these instruments achieved arcminute accuracy,
close to the theoretical resolving power of the human eye.

Though Tycho guarded his intellectual property closely in his lifetime,
knowledge of such instruments spread rapidly after his death in 1601, as his
former assistants received appointments at observatories across Europe and
Asia. As this was happening, Tycho's observations of planetary positions were
being analysed by one of his theoretically-minded former assistants, Johannes
Kepler, who found that the path followed by Mars could not be reproduced by
either Ptolemaic planetary theory or Tycho's proposed replacement for it.
Motivated by this, Kepler went on to develop his own planetary theory, showing
that Tycho's data could be reproduced if the planets followed not circular, but
elliptical orbits about the Sun. This conclusion vindicated Tycho's campaign of
precise observation by demonstrating, as Tycho had hoped to do, that precise
measurements of planetary positions could challenge ancient planetary theory,
even though Kepler had at the same time disproved Tycho's own cosmological
ideas.

Once the case for precision observation had been made, the astrolabe, a small
handheld instrument, was no longer sufficient for astronomers' needs. By the
mid-eighteenth century, the staple instrument for positional astronomy would be
the transit instrument -- an evolved form of Tycho's mural quadrant with a
telescopic sight. Even outside observatories, the astrolabe was by now becoming
largely redundant: as a chronometer, it could no longer compete with the
clockwork rivals which were increasingly widely available and reliable.

\section*{Acknowledgments}

The model presented in this paper came about initially at the suggestion of
Katie Birkwood, Hoyle Project Associate at the Library, St John's College,
Cambridge, who wanted a model astrolabe to use as the focus of an exhibition,
{\it The Way to the Stars: Build Your Own Astrolabe}, during the Cambridge
Science Festival in 2010 March. The model used in that exhibition, and still
available from the associated
website,\footnote{\url{http://www.joh.cam.ac.uk/library/library_exhibitions/schoolresources/astrolabe}}
was a simplified version of that described here. I am grateful to Matthew Smith
for compiling the list of saints' days shown in Figure~\ref{fig:1}, and to
Michael Hoskin for his encouragement and for bringing John North's history of
astronomy to my attention. Finally I would like to thank the referees of this
paper, Mike Frost and Peter Meadows, for making several useful suggestions.


\begin{thebibliography}{}

\bibitem[\protect\citeauthoryear{Chaucer}{Chaucer}{1391}]{r4}Chaucer, G., \textit{Treatise on the Astrolabe}, in {\it The Riverside Chaucer}, ed.\ L.D.\ Benson (Boston, 1987)
\bibitem[\protect\citeauthoryear{Christianson}{Christianson}{2000}]{r13}Christianson J.\ R., {\it On Tycho's Island: Tycho Brahe and his Assistants, 1570--1601}, Cambridge University Press, 2000
\bibitem[\protect\citeauthoryear{Eisner}{Eisner}{1975}]{r1}Eisner, S., \textit{J.\ Brit.\ astr.\ Ass.}, \textbf{86}(1), 18-29 (1975)
\bibitem[\protect\citeauthoryear{Eisner}{Eisner}{1976a}]{r2}Eisner, S., \textit{ibid.}, \textbf{86}(2), 125-132 (1976a)
\bibitem[\protect\citeauthoryear{Eisner}{Eisner}{1976b}]{r3}Eisner, S., \textit{ibid.}, \textbf{86}(3), 219-227 (1976b)
\bibitem[\protect\citeauthoryear{Hoffleit}{Hoffleit}{1964}]{r11}Hoffleit D., {\it Catalogue of Bright Stars}, 3rd rev.ed., Yale University Observatory, 1964
\bibitem[\protect\citeauthoryear{North}{North}{2008}]{r7}North J.D., {\it Cosmos}, University of Chicago Press, 2008
\bibitem[\protect\citeauthoryear{Stahlman \& Gingerich}{Stahlman \& Gingerich}{1963}]{r10}Stahlman W.D.\ \& Gingerich O., {\it Solar and Planetary Longitudes for Years $-2500$ to $+2000$ by 10-day intervals}, University of Wisconsin Press, 1963
\bibitem[\protect\citeauthoryear{Thoren}{Thoren}{1990}]{r12}Thoren V.\ E., {\it The Lord of Uraniborg: A Biography of Tycho Brahe}, Cambridge University Press, 1990
\bibitem[\protect\citeauthoryear{Tuckerman}{Tuckerman}{1962}]{r8}Tuckerman B., {\it Mem.\ American Philosophical Society}, {\bf 56} (1962)
\bibitem[\protect\citeauthoryear{Tuckerman}{Tuckerman}{1964}]{r9}Tuckerman B., {\it ibid.}, {\bf 59} (1964)

\end{thebibliography}
\end{document}